\documentclass[sigconf]{aamas} 

\usepackage{balance}
\usepackage{tabularx}
\usepackage{amsmath}
\usepackage{algorithm}
\usepackage{algpseudocode}
\usepackage{booktabs}
\usepackage{multirow}
\usepackage{makecell}
\usepackage{threeparttable}
\usepackage{tikz}
\usepackage{array}
\usetikzlibrary{decorations.pathreplacing, positioning}

\renewcommand\footnotetextcopyrightpermission[1]{}
\title[Solving Distributed Flexible Job Shop Scheduling Problem Using SAT]{Solving Distributed Flexible Job Shop Scheduling Problem\\ Using SAT}

\acmConference[]{}{}{}

\author{Phuong Dang Anh}
\affiliation{
  \institution{Faculty of Information Technology, VNU University of Engineering and Technology}
  \city{Hanoi}
  \country{Vietnam}}
\email{24020277@vnu.edu.vn}

\author{Duc Nguyen Kim Trung}
\affiliation{
  \institution{Faculty of Information Technology, VNU University of Engineering and Technology}
  \city{Hanoi}
  \country{Vietnam}}
\email{23021533@vnu.edu.vn}

\author{To Van Khanh}
\correspondingauthor
\affiliation{
  \institution{Faculty of Information Technology, VNU University of Engineering and Technology}
  \city{Hanoi}
  \country{Vietnam}}
\email{khanhtv@vnu.edu.vn}

\begin{abstract}
The Distributed Flexible Job Shop Scheduling Problem (DFJSP) requires the joint determination of factory assignment, machine selection, and operation sequencing to minimize the makespan. This paper presents an exact SAT-based approach built on an efficient SAT formulation that exploits problem structure through propagation, symmetry breaking, and incremental SAT solving to effectively address these coupled decisions. The proposed method is evaluated on 76 standard benchmark instances from the Fattahi, Brandimarte, and \textit{rdata} datasets under two- and three-factory configurations. Together with previously established results, the new SAT optimality certificates bring the total number of proven-optimal instances to 72 out of 76, including three newly certified instances; for two of these instances, SAT also improves the previous best-known makespans. Across the benchmark configurations, SAT achieves the best results on four of the five configurations and remains competitive with strong exact approaches such as CP and LBBD1\_CPIL, as well as existing metaheuristic methods. These results demonstrate that the proposed SAT framework is an effective exact approach for the DFJSP, combining strong solution quality with rigorous optimality certification and offering a complementary exact method for future DFJSP studies and evaluations on additional benchmark instances.
\end{abstract}


\newcommand{\BibTeX}{\rm B\kern-.05em{\sc i\kern-.025em b}\kern-.08em\TeX}

\begin{document}

\pagestyle{fancy}
\fancyhead{}

\maketitle 

\section{Introduction}

Production scheduling determines how operations are assigned to manufacturing resources and ordered over time while satisfying technological and capacity constraints. As manufacturing systems become more flexible, these decisions become more difficult because an operation may be processed by different eligible resources rather than by a single predefined machine \cite{dauzereperes2024fjsp}.

The Job Shop Scheduling Problem (JSSP) \cite{graham1966bounds} represents the classical setting in which each operation must be processed on a predefined machine. The Flexible Job Shop Scheduling Problem (FJSP) \cite{brucker1990job} extends this setting by allowing each operation to be processed on one of several eligible machines. Solving the FJSP therefore requires machine selection to be considered together with operation sequencing, which increases the difficulty of the scheduling problem \cite{brandimarte1993routing,fattahi2007mathematical,hurink1994tabu,mastrolilli2000effective}. The Distributed Flexible Job Shop Scheduling Problem (DFJSP), introduced by \citet{degiovanni2010improved}, further extends the FJSP to a multi-factory setting. In the DFJSP, each job is assigned to a single factory, and all of its operations are processed within that factory. A feasible schedule must therefore determine the factory assigned to each job, select an eligible machine for each operation, and determine the processing order of operations competing for the same machine \cite{degiovanni2010improved,meng2020milp}.

A wide range of heuristic and metaheuristic methods has been proposed for the DFJSP, including genetic algorithms, chemical reaction optimization, and hybrid metaheuristics \cite{degiovanni2010improved,chang2017optimisation,lu2018genetic,marzouki2018chemical,xie2023hybrid}. These methods can produce high-quality schedules for relatively large instances and have improved existing benchmark solutions. However, they do not in general provide certificates of optimality or rigorous lower bounds. Consequently, when a good feasible solution is obtained, it may remain unknown whether a better schedule exists. Exact methods address this limitation by providing optimality guarantees. \citet{meng2020milp} developed several Mixed-Integer Linear Programming (MILP) formulations and a Constraint Programming (CP) model for the DFJSP, while \citet{xiong2026lbBD} later proposed Logic-based Benders Decomposition approaches that separate factory assignment from scheduling decisions. These methods have substantially advanced the exact solution of DFJSP benchmark instances. Nevertheless, proving optimality remains difficult for some medium- and large-scale cases within practical computational limits \cite{meng2020milp,xiong2026lbBD}.

These observations motivate the investigation of alternative exact approaches with two complementary capabilities. First, an exact method should be able to determine whether a known solution is optimal and, when possible, provide a proof of optimality. Second, it should remain effective in searching for improved feasible solutions when the current best-known makespan has not yet been certified. A method combining these capabilities is particularly valuable for benchmark-driven research because it can both enlarge the set of instances with proven optimal solutions and improve the best-known results for unresolved cases.

Boolean Satisfiability (SAT) provides an alternative exact framework for discrete optimization. Modern Conflict-Driven Clause Learning SAT solvers combine Boolean propagation, clause learning, conflict analysis, and adaptive search strategies to efficiently solve large combinatorial problems \cite{biere2009handbook}. SAT-based formulations have also been successfully investigated for scheduling and related optimization problems \cite{koshimura2010solving,huang2018efficient,coelho2025comparing}. In particular, previous studies on the classical JSSP have shown that SAT can effectively represent temporal and machine-conflict decisions while supporting makespan minimization through repeated feasibility tests \cite{koshimura2010solving,huang2018efficient}. The DFJSP, however, introduces additional decisions that are absent from the classical JSSP, most notably job-to-factory assignment and alternative machine selection. These decisions interact with operation timing and resource conflicts and introduce additional symmetry when factories are identical. To the best of our knowledge, a dedicated SAT-based exact formulation for the DFJSP has not previously been studied. This work addresses this gap by developing a SAT-based approach for the minimum-makespan DFJSP and evaluating its ability both to certify optimality and to improve previously reported benchmark solutions.

The main contributions of this work are as follows:
\begin{itemize}
\item We propose an exact SAT-based approach for the DFJSP that exploits problem structure through an efficient formulation, problem-specific propagation, symmetry breaking, and incremental SAT solving.

\item We provide a systematic computational evaluation on 76 standard benchmark instances from the Fattahi, Brandimarte \cite{brandimarte1993routing}, and \textit{rdata} \cite{hurink1994tabu} datasets. The results establish three new optimality certificates, two of which correspond to instances for which SAT improves the previous best-known makespans, and, together with previously established results, show that 72 of the 76 instances have now been proven optimal.

\item We provide an additional exact approach for the DFJSP that offers competitive solution quality and rigorous optimality certification, serving as a reference for future studies and evaluations on additional benchmark instances.
\end{itemize}

The remainder of this paper is organized as follows. Section~\ref{sec:problem_literature} defines the DFJSP and reviews related work. Section~\ref{sec:sat_encoding} presents the proposed SAT formulation and the incremental optimization procedure. Section~\ref{sec:computational_experiments} describes the benchmark instances, experimental setup, and computational results. Finally, Section~\ref{sec:conclusion} concludes the paper and outlines directions for future research.

\section{Problem Definition and Related Work}
\label{sec:problem_literature}
\subsection{Problem Definition}
The Distributed Flexible Job Shop Scheduling Problem (DFJSP) consists of a set of independent jobs
$J = \{1, 2, \ldots, n\}$ to be processed in a set of factories $F = \{1, 2, \ldots, f\}$.
All factories are assumed to be identical, meaning that they have the same machine configuration and that the eligibility and processing time of an operation on a given machine do not depend on the factory. Each factory contains the same set of machines $M = \{1, 2, \ldots, m\}$, and its production environment can therefore be viewed as a flexible job shop.

Each job $j \in J$ consists of an ordered sequence of $n_j$ operations, $O_j = \{O_{j,1}, O_{j,2}, \ldots, O_{j,n_j}\}$, which must be processed according to a predefined technological order. Operation $O_{j,r}$ must therefore be completed before operation $O_{j,r+1}$ can start. Each operation $O_{j,r}$ can be processed on one machine selected from a predefined set of eligible machines $M_{j,r} \subseteq M$. The processing time of operation $O_{j,r}$ on an eligible machine $k \in M_{j,r}$ is denoted by $p_{j,r,k}$.

Each job is assigned to exactly one factory, and all operations belonging to that job must be processed within the assigned factory. Within each factory, every operation is assigned to exactly one eligible machine, and operations assigned to the same machine must be processed without overlap.

The DFJSP considered in this work is defined under deterministic scheduling assumptions. All jobs, factories, and machines are available at time zero. Each machine can process at most one operation at a time, and operation processing is non-preemptive. Processing times are deterministic and known in advance. Machine breakdowns and setup times are not considered.

The objective is to minimize the makespan, defined as the completion time of the last finished job in the schedule $C_{\max} = \max_{j \in J} C_j$, where $C_j$ denotes the completion time of job $j$. The optimal makespan is subsequently denoted by $C_{\max}^* = \min C_{\max}$.

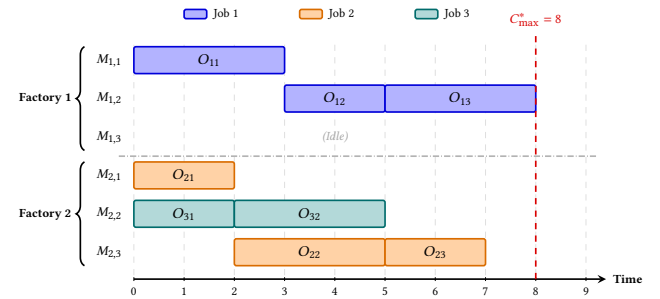
\begin{figure}[htbp]
  \centering
  \resizebox{\columnwidth}{!}{%
  \begin{tikzpicture}[x=0.85cm, y=0.65cm, >=stealth]

    \tikzset{
      job1/.style={fill=blue!30, draw=blue!85!black, thick, rounded corners=1pt},
      job2/.style={fill=orange!35, draw=orange!85!black, thick, rounded corners=1pt},
      job3/.style={fill=teal!30, draw=teal!85!black, thick, rounded corners=1pt},
      gridline/.style={draw=gray!25, dashed, line width=0.4pt}
    }

    \foreach \t in {0, 1, ..., 9} {
      \draw[gridline] (\t, -0.6) -- (\t, 5.5);
    }

    \draw[densely dashdotted, gray!60, thick] (-0.3, 2.5) -- (9.2, 2.5);

    \draw[job1] (0, 4.65) rectangle (3, 5.35) node[midway, font=\footnotesize\bfseries] {$O_{11}$};
    \draw[job1] (3, 3.65) rectangle (5, 4.35) node[midway, font=\footnotesize\bfseries] {$O_{12}$};
    \draw[job1] (5, 3.65) rectangle (8, 4.35) node[midway, font=\footnotesize\bfseries] {$O_{13}$};
    \node[font=\scriptsize\itshape, text=gray!70] at (4, 3.0) {(Idle)};

    \draw[job2] (0, 1.65) rectangle (2, 2.35) node[midway, font=\footnotesize\bfseries] {$O_{21}$};

    \draw[job3] (0, 0.65) rectangle (2, 1.35) node[midway, font=\footnotesize\bfseries] {$O_{31}$};
    \draw[job3] (2, 0.65) rectangle (5, 1.35) node[midway, font=\footnotesize\bfseries] {$O_{32}$};

    \draw[job2] (2, -0.35) rectangle (5, 0.35) node[midway, font=\footnotesize\bfseries] {$O_{22}$};
    \draw[job2] (5, -0.35) rectangle (7, 0.35) node[midway, font=\footnotesize\bfseries] {$O_{23}$};

    \draw[->, thick] (0, -0.7) -- (9.4, -0.7) node[right, font=\scriptsize\bfseries] {Time};
    \foreach \t in {0, 1, ..., 9} {
      \draw (\t, -0.65) -- (\t, -0.75) node[below, font=\scriptsize] {\t};
    }

    \node[anchor=east, font=\scriptsize\bfseries] at (-0.1, 5.0) {$M_{1,1}$};
    \node[anchor=east, font=\scriptsize\bfseries] at (-0.1, 4.0) {$M_{1,2}$};
    \node[anchor=east, font=\scriptsize\bfseries] at (-0.1, 3.0) {$M_{1,3}$};
    \node[anchor=east, font=\scriptsize\bfseries] at (-0.1, 2.0) {$M_{2,1}$};
    \node[anchor=east, font=\scriptsize\bfseries] at (-0.1, 1.0) {$M_{2,2}$};
    \node[anchor=east, font=\scriptsize\bfseries] at (-0.1, 0.0) {$M_{2,3}$};

    \draw[decorate, decoration={brace, amplitude=4pt, raise=2pt}, thick] 
      (-0.9, 2.65) -- (-0.9, 5.35) node[midway, left=5pt, font=\scriptsize\bfseries, align=center] {Factory 1};
    \draw[decorate, decoration={brace, amplitude=4pt, raise=2pt}, thick] 
      (-0.9, -0.35) -- (-0.9, 2.35) node[midway, left=5pt, font=\scriptsize\bfseries, align=center] {Factory 2};

    \draw[red!85!black, thick, dashed] (8, -0.7) -- (8, 5.7) 
      node[above, font=\scriptsize\bfseries, text=red!85!black] {$C_{\max}^* = 8$};

    \begin{scope}[shift={(1.0, 6.2)}]
      \draw[job1] (0, -0.12) rectangle (0.45, 0.12);
      \node[right, font=\scriptsize] at (0.45, 0) {Job 1};

      \draw[job2] (2.3, -0.12) rectangle (2.75, 0.12);
      \node[right, font=\scriptsize] at (2.75, 0) {Job 2};

      \draw[job3] (4.6, -0.12) rectangle (5.05, 0.12);
      \node[right, font=\scriptsize] at (5.05, 0) {Job 3};
    \end{scope}

  \end{tikzpicture}
  }
  \caption{Optimal DFJSP schedule.}
  \label{fig:optimal_dfjsp_homogeneous}
\end{figure}

Table~\ref{tab:dfjsp_homogeneous} presents a small DFJSP instance consisting of three jobs, two identical factories, and three machines in each factory. Job $J_1$ and job $J_2$ each contain three operations, whereas job $J_3$ contains two operations. Each operation can be processed only on its eligible machines, with the corresponding processing times given in the table. Since the factories are identical, the eligibility and processing times shown in the table apply equally to every factory. Figure~\ref{fig:optimal_dfjsp_homogeneous} illustrates the optimal schedule with $C_{\max}^* = 8$.

\begin{table}[htbp]
  \centering
  \caption{Illustrative DFJSP instance.}
  \label{tab:dfjsp_homogeneous}
  \begin{tabular}{cc *{3}{w{c}{1.2cm}}}
    \toprule
    \multirow{2}{*}{Job} & \multirow{2}{*}{Operation} & \multicolumn{3}{c}{Machine processing time} \\
    \cmidrule(lr){3-5}
    & & $M_1$ & $M_2$ & $M_3$ \\
    \midrule
    \multirow{3}{*}{$J_1$} 
      & $O_{11}$ & 3   & 4   & --  \\
      & $O_{12}$ & --  & 2   & 3   \\
      & $O_{13}$ & 4   & 3   & 5   \\
    \midrule
    \multirow{3}{*}{$J_2$} 
      & $O_{21}$ & 2   & 3   & --  \\
      & $O_{22}$ & 4   & --  & 3   \\
      & $O_{23}$ & --  & 3   & 2   \\
    \midrule
    \multirow{2}{*}{$J_3$} 
      & $O_{31}$ & 3   & 2   & 4   \\
      & $O_{32}$ & 4   & 3   & --  \\
    \bottomrule
  \end{tabular}
\end{table}

\subsection{Related Work}
\label{sec:literature_review}
Research on the DFJSP has mainly focused on two broad classes of solution approaches: heuristic and metaheuristic methods for efficiently obtaining feasible schedules, and exact methods aimed at providing optimality guarantees. More recently, advances in exact optimization have also led to decomposition-based approaches that exploit the hierarchical structure of factory assignment and shop-level scheduling.

Heuristic and metaheuristic approaches have been extensively investigated for the DFJSP because of the large search space arising from factory assignment, machine selection, and operation sequencing. Representative approaches include genetic algorithms, chemical reaction optimization, and other problem-specific search procedures \cite{degiovanni2010improved,chang2017optimisation,lu2018genetic,marzouki2018chemical}. More recently, \citet{xie2023hybrid} proposed a Hybrid Genetic Tabu Search Algorithm (HGTSA) that combines genetic operators with critical-path-based tabu search and reported 13 new upper bounds on 69 benchmark instances. These approaches have played an important role in the search for competitive feasible schedules and in improving known benchmark upper bounds. However, heuristic and metaheuristic methods do not generally provide valid lower bounds or certificates of optimality.

Exact approaches complement heuristic search by providing mechanisms for establishing optimality. \citet{meng2020milp} developed four Mixed-Integer Linear Programming (MILP) formulations and a Constraint Programming (CP) model for the DFJSP. Their computational results showed that the CP model scales more effectively than the MILP formulations on larger instances. The study established optimal solutions for 62 benchmark instances and reported 11 new best solutions, showing that exact methods can contribute both to optimality certification and to the improvement of known upper bounds. More recently, \citet{xiong2026lbBD} proposed Logic-based Benders Decomposition (LBBD) approaches that separate job-to-factory assignment from the scheduling subproblems associated with individual factories. Their best-performing variant, LBBD1\_CPIL, combines a master assignment problem with CP-based scheduling and reported an average optimality gap below 1.2\% over 286 benchmark instances. These results substantially extended the range of DFJSP instances that can be addressed within an exact optimization framework, although proving optimality remains difficult for some medium- and large-scale instances.

Boolean Satisfiability (SAT) provides another exact paradigm for combinatorial scheduling and has previously been investigated for the classical Job Shop Scheduling Problem. \citet{koshimura2010solving} formulated makespan minimization as a sequence of SAT feasibility problems using Boolean variables to represent temporal and machine-ordering decisions. Their approach proved optimal makespans for previously unresolved benchmark instances and improved several known bounds. \citet{huang2018efficient} subsequently developed a more compact SAT formulation by restricting temporal variables and clauses to feasible time ranges and introducing precedence variables only for operations that can compete for the same machine. These studies indicate that SAT can be effective for exact scheduling when the Boolean formulation exploits problem-specific structure to strengthen propagation and avoid unnecessary variables and clauses.

However, SAT formulations developed for the classical JSSP do not directly capture the additional decisions introduced by the DFJSP. In the JSSP, each operation is assigned to a predefined machine and all operations are processed within a single shop. In the DFJSP, each job must additionally be assigned to a factory, while each operation must select one of several eligible machines. Consequently, machine-capacity conflicts are conditional on both factory and machine assignments: two operations compete for the same physical resource only when their jobs are assigned to the same factory and both operations select the same machine. Moreover, identical factories introduce symmetries associated with interchangeable factory labels. These characteristics require factory assignment, machine selection, temporal decisions, and conditional resource conflicts to be represented jointly within the Boolean formulation.

Existing exact approaches for the DFJSP have primarily relied on MILP, CP, and decomposition-based methods, while SAT-based scheduling studies have addressed various scheduling settings. The proposed SAT formulation jointly models factory assignment, machine selection, timing, precedence, and conditional machine-capacity constraints, supported by temporal preprocessing, propagation, symmetry breaking, and incremental SAT solving.

\section{SAT Solving for the Distributed Flexible Job Shop Scheduling Problem}
\label{sec:sat_encoding}

For the formulation, all operations are indexed globally by $i\in O$, and $J(i)$ denotes the job containing operation $i$. Let $E$ denote the set of direct precedence relations, $M_i$ the set of eligible machines for operation $i$, and $p_{i,k}$ its processing time on machine $k\in M_i$. 

\subsection{Upper-Bound Construction and Preprocessing}

A finite makespan bound is required to construct the time-indexed SAT encoding because it determines the temporal range over which start-time variables are instantiated. A smaller valid upper bound therefore reduces the number of Boolean variables and clauses and provides the initial bound for the subsequent makespan optimization.

Before constructing the SAT model, we compute an initial feasible schedule to obtain an upper bound $UB$ on the makespan. Let $p_i^{\min}=\min_{k\in M_i}p_{i,k}$ denote the minimum processing time of operation $i$. Since the operations of each job form a linear precedence chain, let $\operatorname{pred}(i)$ and $\operatorname{succ}(i)$ denote the immediate predecessor and successor of operation $i$, respectively, whenever they exist. The workload of job $j$ is estimated as
\begin{equation}
 W_j=\sum_{i:J(i)=j}p_i^{\min}.   
\end{equation}

Jobs are first sorted in non-increasing order of $W_j$ and assigned sequentially to the factory with the smallest accumulated workload. Once the factory assignments are fixed, the operations are scheduled greedily. A tail length is computed for each operation according to
\begin{equation}
L_i =
\begin{cases}
p_i^{\min}, & \text{if } i \text{ has no successor},\\
p_i^{\min}+L_{\operatorname{succ}(i)}, & \text{otherwise}.
\end{cases}
\label{eq:tail_length}
\end{equation}

Among the currently available operations, the operation with the largest tail length is selected. For each eligible machine $k$ in the assigned factory, the operation is scheduled at the maximum of its precedence-based ready time and the current availability time of machine $k$. The machine yielding the smallest resulting completion time is selected. The makespan of the resulting feasible schedule is used as $UB$.

Using $UB$, precedence-based temporal bounds are then computed for each operation. The earliest start times are obtained by forward propagation:
\begin{equation}
ES_i =
\begin{cases}
0, & \text{if } i \text{ is the first operation of its job},\\
ES_{\operatorname{pred}(i)} + p_{\operatorname{pred}(i)}^{\min}, & \text{otherwise}.
\end{cases}
\label{eq:earliest_start}
\end{equation}

The latest start times are initialized from $UB$ and tightened by backward propagation:
\begin{equation}
LS_i =
\begin{cases}
UB-p_i^{\min}, & \text{if } i \text{ is the last operation of its job},\\
LS_{\operatorname{succ}(i)}-p_i^{\min}, & \text{otherwise}.
\end{cases}
\label{eq:latest_start}
\end{equation}

The resulting interval $[ES_i,LS_i]$ defines the temporal domain of operation $i$ and is used to restrict the time-indexed variables in the SAT formulation.

\subsection{SAT Formulation}
\label{sec:sat_formulation}

Using the temporal domains obtained above, we construct the SAT formulation. The eligible machines of each operation are ordered by non-decreasing processing time. Let $M_i'=(k_{i,1},k_{i,2},\ldots,k_{i,\lvert M_i\rvert})$ denote this ordering, such that
$p_{i,k_{i,1}}\leq p_{i,k_{i,2}}\leq\cdots\leq p_{i,k_{i,\lvert M_i\rvert}}$.

\paragraph{Boolean variables.}

The Boolean variables used in the formulation are summarized in Table~\ref{tab:sat_variables}.
\begin{table}[htbp]
\centering
\caption{Boolean variables used in the SAT formulation.}
\label{tab:sat_variables}
\begin{tabular}{>{\raggedright\arraybackslash}p{2.0cm} >{\raggedright\arraybackslash}p{5.1cm}}
\toprule
Variable & Meaning \\
\midrule
$f_{j,q}$ & True iff job $j$ is assigned to factory $q$. \\
$sf_{j,\ell}$ & True iff jobs $j$ and $\ell$ are assigned to the same factory. \\
$m_{i,k}$ & True iff operation $i$ is processed on machine $k$. \\
$xm_{i,r}$ & True iff the machine selected for operation $i$ is at position $r$ or later in the processing-time-ordered list $M_i'$. \\
$s_{i,t}$ & True iff operation $i$ starts at time $t$. \\
$x_{i,t}$ & True iff operation $i$ starts at or after time $t$. \\
$sm_{i,h,k}$ & True iff operations $i$ and $h$ both select machine $k$. \\
$sfm_{i,h,k}$ & True iff operations $i$ and $h$ use machine $k$ in the same factory. \\
\bottomrule
\end{tabular}
\end{table}

\paragraph{Factory assignment constraints.}

Since all factories are identical, permuting their indices does not change the feasibility or makespan of a schedule. We therefore restrict the factories available to job $j$ to $F_j=\{1,\ldots,\min(j,f)\}$. Thus, the first job can only be assigned to factory 1, the second job can use factories 1 and 2, and so on. Under this restricted domain, each job is assigned to exactly one factory:
\begin{equation}
\sum_{q\in F_j} f_{j,q}=1,
\qquad
\forall j\in J.
\label{eq:factory_exactly_one}
\end{equation}

\paragraph{Machine assignment constraints.}
The machine-selection variables are encoded using the processing-time ordering of $M_i'$. Specifically, the order variables form a monotone sequence, and the selected machine is identified by the unique transition from true to false in this sequence:
\begin{equation}
\begin{aligned}
& xm_{i,1} = \text{True}, \\
& xm_{i,r+1} \rightarrow xm_{i,r}, && r = 1, \ldots, \lvert M_i\rvert - 1, \\
& m_{i,k_{i,r}} \leftrightarrow \left(xm_{i,r}\land\neg xm_{i,r+1}\right), \qquad && r = 1, \ldots, \lvert M_i\rvert - 1, \\
& m_{i,k_{i,\lvert M_i\rvert}} \leftrightarrow xm_{i,\lvert M_i\rvert}.
\end{aligned}
\label{eq:machine_assignment}
\end{equation}

\paragraph{Start-time encoding constraints.}
Similarly to the machine-assignment encoding, the start time of each operation $i$ is encoded over its temporal domain $[ES_i,LS_i]$ using order variables:
\begin{equation}
\begin{aligned}
& x_{i,ES_i} = \text{True}, \\
& x_{i,t+1} \rightarrow x_{i,t}, && t = ES_i, \ldots, LS_i - 1, \\
& s_{i,t} \leftrightarrow \left(x_{i,t}\land\neg x_{i,t+1}\right), \qquad && t = ES_i, \ldots, LS_i - 1, \\
& s_{i,LS_i} \leftrightarrow x_{i,LS_i}.
\end{aligned}
\label{eq:start_time_encoding}
\end{equation}

\paragraph{Direct precedence constraints.}

For each precedence relation $(i,h)\in E$ and start time $t\in[ES_i,LS_i]$, we first establish a separation based on the minimum processing time $p_i^{\min}$:
\begin{equation}
\begin{aligned}
& s_{i,t}\rightarrow x_{h,t+p_i^{\min}}, && ES_h < t+p_i^{\min}\leq LS_h, \\
& \neg s_{i,t}, && t+p_i^{\min}>LS_h.
\end{aligned}
\label{eq:precedence_baseline}
\end{equation}

To account for slower machine alternatives, this separation is conditionally tightened for each position $r=2,\ldots,\lvert M_i\rvert$ satisfying $p_{i,k_{i,r}}>p_{i,k_{i,r-1}}$:
\begin{equation}
\begin{aligned}
& s_{i,t}\land xm_{i,r} \rightarrow x_{h,t+p_{i,k_{i,r}}}, \qquad && ES_h < t+p_{i,k_{i,r}}\leq LS_h, \\
& s_{i,t}\rightarrow\neg xm_{i,r}, && t+p_{i,k_{i,r}}>LS_h.
\end{aligned}
\label{eq:precedence_strengthened}
\end{equation}

In both cases, instances with $t+p\leq ES_h$ are trivially satisfied by the time window of $h$ and omitted.

\paragraph{Transitive precedence constraints.}
Let $E^*$ denote the transitive closure of $E$. Because the operations of each job form a linear chain, there exists a unique precedence path $P(i,h)$ from operation $i$ to any downstream operation $h$, with a cumulative minimum processing time $w_{i,h}=\sum_{v\in P(i,h)\setminus\{h\}}p_v^{\min}$. For each non-direct relation $(i,h)\in E^*\setminus E$ and each start time $t\in[ES_i,LS_i]$, we enforce:
\begin{equation}
\begin{aligned}
& s_{i,t}\rightarrow x_{h,t+w_{i,h}}, \qquad && ES_h < t+w_{i,h}\leq LS_h, \\
& \neg s_{i,t}, && t+w_{i,h}>LS_h.
\end{aligned}
\label{eq:transitive_precedence}
\end{equation}
Cases where $t+w_{i,h}\leq ES_h$ are inherently satisfied by the temporal domain of $h$ and omitted. While these constraints do not alter the set of feasible schedules, they enable the temporal implications of composite precedence chains to be propagated directly via unit propagation.

\paragraph{Machine-capacity constraints.}
Because each machine can process at most one operation at a time, machine-capacity constraints must ensure that operations sharing the same machine within the same factory do not overlap in time. These constraints are required only for unordered pairs of operations $\{i, h\} \subseteq O$ with $(i,h), (h,i)\notin E^*$ that share an eligible machine $k\in M_i\cap M_h$.

Let $j=\min\{J(i),J(h)\}$ and $\ell=\max\{J(i),J(h)\}$. Given that $F_j\subseteq F_\ell$ for $j<\ell$, the conditions that the two operations are assigned to the same factory, select the same machine, and consequently use the same physical resource are formulated as:
\begin{equation}
\begin{aligned}
& f_{j,q}\land f_{\ell,q} \rightarrow sf_{j,\ell}, && q\in F_j, \\
& sf_{j,\ell}\land f_{j,q} \rightarrow f_{\ell,q}, && q\in F_j, \\
& sm_{i,h,k} \leftrightarrow \left(m_{i,k}\land m_{h,k}\right), \\
& sfm_{i,h,k} \leftrightarrow \left(sm_{i,h,k}\land sf_{j,\ell}\right).
\end{aligned}
\label{eq:resource_sharing}
\end{equation}

The machine capacity limit dictates that operations $i$ and $h$ cannot be executed concurrently on machine $k$. We enforce this capacity constraint by requiring that if both operations are assigned to machine $k$ and operation $i$ starts at time $t$, operation $h$ must either complete before $i$ starts or start after $i$ completes. For each $t \in [ES_i, LS_i]$, letting $A = t - p_{h,k}$ and $B = t + p_{i,k}$, this requirement is encoded as:
\begin{equation}
sfm_{i,h,k} \land s_{i,t} \to (\neg x_{h,A+1} \lor x_{h,B}),
\label{eq:machine_capacity}
\end{equation}
where literal $\neg x_{h,A+1}$ is included only when $A \geq ES_h$, and $x_{h,B}$ is included only when $B \leq LS_h$. Instances of Equation~\eqref{eq:machine_capacity} with $A \geq LS_h$ or $B \leq ES_h$ are trivially satisfied, and thus no clause is generated. Conversely, when both $A < ES_h$ and $B > LS_h$, neither literal is included on the right-hand side, resulting in a 2-literal clause $\neg sfm_{i,h,k} \lor \neg s_{i,t}$ (i.e., $sfm_{i,h,k} \land s_{i,t} \to \bot$), which forbids operation $i$ from starting at time $t$ whenever the two operations share machine $k$ in the same factory.

\paragraph{Symmetry breaking.}
The restricted factory domains $F_j$ eliminate some permutations of identical factories but do not remove all equivalent assignments. To break the remaining symmetries, we impose the condition that factory $q$ can be used only if factory $q-1$ has already been assigned to an earlier job:
\begin{equation}
f_{j,q}
\rightarrow
\bigvee_{h=q-1}^{j-1} f_{h,q-1},
\qquad
j=2,\ldots,\lvert J\rvert,
\quad
q\in F_j\setminus\{1\}.
\label{eq:factory_symmetry}
\end{equation}

\subsection{Incremental SAT Solving}
\label{sec:incremental}

Because SAT is fundamentally a decision problem, SAT solvers cannot directly optimize an objective function. Therefore, makespan minimization must be achieved by solving a sequence of decision problems, each querying whether a feasible schedule exists within a candidate makespan bound $U$. 

To enforce a specific candidate bound $U$ in the SAT formulation, it is sufficient to restrict the completion times of the last operations of all jobs. Let $O^{\mathrm{last}}$ denote the set of these final operations. For each $i\in O^{\mathrm{last}}$ and each eligible machine $k\in M_i$, we define the machine-dependent latest start time:
\begin{equation}
\tau_{i,k}(U)=U-p_{i,k}.
\label{eq:makespan_start_limit}
\end{equation}

If operation $i$ is assigned to machine $k$, it must start no later than $\tau_{i,k}(U)$ to ensure the entire job completes within the current makespan bound. The corresponding constraints are:
\begin{equation}
\begin{aligned}
&\neg m_{i,k},
&& \tau_{i,k}(U)<ES_i,\\
&m_{i,k}\rightarrow\neg x_{i,\tau_{i,k}(U)+1},
&& ES_i\leq\tau_{i,k}(U)<LS_i.
\end{aligned}
\label{eq:makespan_bound}
\end{equation}

When $\tau_{i,k}(U)\geq LS_i$, no additional clause is required because the temporal domain of operation $i$ already guarantees its completion within $U$ when machine $k$ is selected.

Algorithm~\ref{alg:incremental_makespan} summarizes the incremental optimization procedure. The base SAT formulation is constructed once using the temporal domains derived from the initial upper bound $UB$. Starting from $UB$, makespan clauses corresponding to the current candidate bound are incrementally added to the same solver instance. Whenever a feasible solution is obtained, its actual makespan is used to tighten the next bound directly.

\begin{algorithm}[htbp]
\caption{Incremental makespan optimization}
\label{alg:incremental_makespan}
\begin{algorithmic}[1]
\Require SAT solver $S$ containing the base formulation, initial upper bound $UB$
\Ensure Optimal schedule and optimal makespan $C_{\max}^{*}$
\State $U \gets UB$
\State $\mathit{bestSchedule} \gets \varnothing$
\State $C_{\max}^{*} \gets UB$
\While{true}
    \State Add the makespan constraints of Eq.~\eqref{eq:makespan_bound} for bound $U$ to $S$
    \State $(\mathit{status}, \mathit{schedule}) \gets \Call{Solve}{S}$
    \If{$\mathit{status} = \mathrm{UNSAT}$}
        \State \Return $(\mathit{bestSchedule}, C_{\max}^{*})$
    \EndIf
    \State $C_{\max} \gets \Call{Makespan}{\mathit{schedule}}$
    \State $\mathit{bestSchedule} \gets \mathit{schedule}$
    \State $C_{\max}^{*} \gets C_{\max}$
    \State $U \gets C_{\max} - 1$
\EndWhile
\end{algorithmic}
\end{algorithm}

The same solver instance is retained throughout the optimization process, so all previously added clauses and learned information remain available in subsequent iterations. After obtaining a feasible schedule with makespan $C_{\max}$, the next bound is set directly to $C_{\max}-1$. Because makespan constraints are added monotonically, all previously imposed upper bounds remain valid and are implied by the tighter bound. When the formulation becomes unsatisfiable, no feasible schedule exists with a makespan smaller than the last feasible $C_{\max}$. Therefore, the last feasible schedule is optimal and its makespan is $C_{\max}^{*}$.

\section{Computational Experiments}
\label{sec:computational_experiments}
\subsection{Benchmark Instances}
We evaluate the proposed SAT method on three widely used FJSP benchmark suites: Brandimarte~\cite{brandimarte1993routing}, Fattahi et al.~\cite{fattahi2007mathematical}, and the \textit{rdata} subset of Hurink et al.~\cite{hurink1994tabu}. Since these benchmarks were originally designed for single-factory production, we use the distributed versions adopted in previous DFJSP studies~\cite{meng2020milp,xie2023hybrid}, in which identical factories share the same machine configurations and processing capabilities.
The Fattahi set contains 10 instances and is evaluated with two factories, while the Brandimarte set contains 10 instances and is tested with both two and three factories. The \textit{rdata} set contains 23 instances and is also evaluated with two and three factories. We omit the four-factory \textit{rdata} setting because previous studies~\cite{meng2020milp,xie2023hybrid} show that almost all instances already attain their theoretical lower bounds with three factories, while the four-factory configuration becomes nearly trivial and therefore provides limited discriminatory value.

\subsection{Experimental Setup and Metrics}
All SAT experiments are conducted on a Google Cloud Compute Engine \textit{c4-standard-4} virtual machine. The proposed encoding is implemented in Python using PySAT~\cite{itk-sat24}, with CaDiCaL version~1.9.5~\cite{BiereFazekasFleuryHeisinger-SAT-Competition-2020-solvers} as the underlying solver. Each instance is evaluated under 600~s and 1800~s time limits, including preprocessing, encoding construction, and solving. The 1800~s run is omitted when the corresponding instance has already been proven optimal within 600~s.

The proposed method is compared with the MILP and CP models of Meng et al.~\cite{meng2020milp}, HGTSA~\cite{xie2023hybrid}, and LBBD1\_CPIL~\cite{xiong2026lbBD}. For the CP model of Meng et al., two sets of results are reported separately, namely the original 600~s results from Meng et al.~\cite{meng2020milp} and the 1800~s results obtained by Xiong et al.~\cite{xiong2026lbBD} using the same model. All baseline results are taken from the corresponding publications or publicly released experimental data, whereas only the proposed SAT method is executed on the platform described above.

HGTSA~\cite{xie2023hybrid} is run for up to 4000 iterations rather than under a fixed time limit. As a metaheuristic, it does not itself prove optimality; therefore, a reported HGTSA solution is regarded as optimal when its makespan matches the corresponding reported lower bound. To ensure consistency with previous literature, we consider only the 20 two-factory \textit{rdata} instances reported by \citet{xiong2026lbBD} (for LBBD1\_CPIL and CP) that match the benchmark set used in earlier studies. The aggregate metrics for LBBD1\_CPIL and the CP rerun are therefore computed over these 20 common instances\footnote{The corresponding data and results are publicly available at \url{https://github.com/FuliXiong/Instances-and-Results-for-DFJSP}}.

For reproducibility, both the source code of the proposed SAT method and the detailed per-instance experimental results are provided in the Supplementary Material accompanying this paper.

We use the following metrics to evaluate solution quality and computational effectiveness:
\begin{itemize}
    \item \textit{\#Optimal}: The number of instances for which the reported makespan is established as optimal. For HGTSA, optimality is inferred when its reported best makespan matches the reported lower bound.
    \item \textit{\#Best}: The number of instances for which a method obtains the lowest makespan among all compared approaches. Ties are credited to all corresponding methods.
    \item \textit{\#New}: The number of instances for which SAT obtains a makespan strictly smaller than the previously best-known published value.
    \item \textit{Avg.\ Dev.\ (\%)}: The average percentage deviation from the best-found makespan over the evaluated instance set~$I$:
    \begin{equation*}
    \text{Avg.\ Dev.\ (\%)} = \frac{1}{|I|} \sum_{i \in I} \frac{C_{\max,i} - C_{\max,i}^{{\text{best}}}}{C_{\max,i}^{\text{best}}} \times 100\%.
    \end{equation*}
    Here, $C_{\max,i}$ denotes the makespan obtained by the evaluated method for instance~$i$, while $C_{\max,i}^{\mathrm{best}}$ denotes the minimum makespan reported among all compared methods for that instance.
\end{itemize}

\subsection{Results and Analysis}

\begin{table*}[htbp]
\centering
\begin{threeparttable}
\caption{Comparison of experimental results across different datasets and methods.}
\label{tab:experimental_summary}
\small
\setlength{\tabcolsep}{3.5pt}
\begin{tabular}{lcclccccccc}
\toprule
\makecell[t]{\textbf{Dataset}} & 
\makecell[t]{\textbf{\#Factories}} & 
\makecell[t]{\textbf{\#Instances}} & 
\makecell[t]{\textbf{Metric}} &
\makecell[t]{\textbf{MILP~\cite{meng2020milp}} \\ \textbf{(600s)}} &
\makecell[t]{\textbf{CP~\cite{meng2020milp}} \\ \textbf{(600s)}} &
\makecell[t]{\textbf{HGTSA~\cite{xie2023hybrid}}} &
\makecell[t]{\textbf{LBBD1\_CPIL~\cite{xiong2026lbBD}} \\ \textbf{(1800s)}} &
\makecell[t]{\textbf{CP~\cite{xiong2026lbBD}} \\ \textbf{(1800s)}} &
\makecell[t]{\textbf{SAT} \\ \textbf{(600s)}} &
\makecell[t]{\textbf{SAT} \\ \textbf{(1800s)}} \\
\midrule
\multirow{8}{*}{Brandimarte~\cite{brandimarte1993routing}}
& \multirow{4}{*}{2} & \multirow{4}{*}{10} & \#Optimal       & 3     & -- & -- & 4    & \textbf{7}    & 6     & \textbf{7} \\
&                    &                     & \#Best          & 3     & -- & -- & 8    & 7             & 8     & \textbf{9} \\
&                    &                     & \#New           & --    & -- & -- & --   & --            & \textbf{1} & \textbf{1} \\
&                    &                     & Avg. Dev. (\%)  & 50.17 & -- & -- & 0.58 & 0.88          & 0.31  & \textbf{0.11} \\
\cline{2-11}
& \multirow{4}{*}{3} & \multirow{4}{*}{10} & \#Optimal       & --    & -- & -- & 5    & 8             & \textbf{10} & -- \\
&                    &                     & \#Best          & --    & -- & -- & 5    & \textbf{10}   & \textbf{10} & -- \\
&                    &                     & \#New           & --    & -- & -- & --   & --            & 0           & -- \\
&                    &                     & Avg. Dev. (\%)  & --    & -- & -- & 1.79 & \textbf{0.00} & \textbf{0.00} & -- \\
\midrule
\multirow{4}{*}{Fattahi~\cite{fattahi2007mathematical}}
& \multirow{4}{*}{2} & \multirow{4}{*}{10} & \#Optimal       & \textbf{10} & -- & -- & -- & -- & \textbf{10} & -- \\
&                    &                     & \#Best          & \textbf{10} & -- & -- & -- & -- & \textbf{10} & -- \\
&                    &                     & \#New           & --          & -- & -- & -- & -- & 0           & -- \\
&                    &                     & Avg. Dev. (\%)  & \textbf{0.00} & -- & -- & -- & -- & \textbf{0.00} & -- \\
\midrule
\multirow{8}{*}{\textit{rdata}~\cite{hurink1994tabu}}
& \multirow{4}{*}{2} & \multirow{4}{*}{23} & \#Optimal       & 13   & \textbf{17} & 14          & 15/20\tnote{*} & 14/20\tnote{*} & \textbf{17} & \textbf{17} \\
&                    &                     & \#Best          & 13   & 17          & \textbf{23} & 16/20\tnote{*} & 14/20\tnote{*} & 17          & 17 \\
&                    &                     & \#New           & --   & --          & --          & --             & --             & 0           & 0 \\
&                    &                     & Avg. Dev. (\%)  & 1.45 & 0.35        & \textbf{0.00} & 0.12\tnote{*} & 0.27\tnote{*} & 0.39        & 0.24 \\
\cline{2-11}
& \multirow{4}{*}{3} & \multirow{4}{*}{23} & \#Optimal       & 20   & 22   & 22   & -- & -- & \textbf{23}   & -- \\
&                    &                     & \#Best          & 20   & 22   & 22   & -- & -- & \textbf{23}   & -- \\
&                    &                     & \#New           & --   & --   & --   & -- & -- & \textbf{1}    & -- \\
&                    &                     & Avg. Dev. (\%)  & 0.55 & 0.01 & 0.09 & -- & -- & \textbf{0.00} & -- \\
\bottomrule
\end{tabular}
\begin{tablenotes}[flushleft]
\footnotesize
\item ``--'' denotes not applicable, unreported results, or omitted runs where SAT (600s) already solved all instances to optimality.
\item[*] Values are computed over the 20 common two-factory \textit{rdata} instances reported by Xiong et al.~\cite{xiong2026lbBD}.
\end{tablenotes}
\end{threeparttable}
\end{table*}

Table~\ref{tab:experimental_summary} summarizes the computational results across the five benchmark configurations. Within 600~s, SAT obtains the best makespan among the compared methods on 68 of the 76 instances and proves optimality for 66 instances. Extending the unresolved runs to 1800~s increases these numbers to 69 and 67, respectively. SAT proves all Fattahi instances optimal, matching the complete result previously reported for MILP, and also solves all three-factory Brandimarte and \textit{rdata} instances to proven optimality within 600~s. The remaining unresolved cases are concentrated in the two-factory Brandimarte and \textit{rdata} configurations, where the smaller number of factories leads to stronger competition for machine resources and more restrictive scheduling decisions. Across the five benchmark configurations, SAT achieves or matches the strongest overall performance in four of them: two-factory Brandimarte, three-factory Brandimarte, two-factory Fattahi, and three-factory \textit{rdata}. The exception is the two-factory \textit{rdata} configuration, where HGTSA obtains the best makespan on all 23 instances, although SAT still proves 17 instances optimal and remains competitive in solution quality.

The two-factory configurations provide the most informative comparison among the exact approaches. On Brandimarte, SAT and the CP rerun of Xiong et al.~\cite{xiong2026lbBD} both prove seven instances optimal within 1800~s. However, SAT obtains the best makespan on nine instances, compared with seven for CP, and achieves an average deviation of 0.11\%, compared with 0.88\% for CP. LBBD1\_CPIL obtains the best makespan on eight instances but proves four instances optimal. On the two-factory \textit{rdata} configuration, the original CP results of Meng et al.~\cite{meng2020milp} and SAT each prove 17 of the 23 instances optimal. HGTSA obtains the best reported makespan on all 23 instances, while SAT achieves 17 best solutions and an average deviation of 0.24\% after 1800~s. LBBD1\_CPIL proves 15 of the 20 reported instances optimal and obtains the best makespan on 16 of them. These results indicate that SAT is competitive not only in optimality certification but also in solution quality on the unresolved instances.

The difference between the two- and three-factory configurations is noticeable for Brandimarte and \textit{rdata}. With three factories, SAT proves all 10 Brandimarte and all 23 \textit{rdata} instances optimal within 600~s. For Brandimarte, the CP rerun reaches the same best makespan on all 10 instances but certifies only eight within 1800~s, while LBBD1\_CPIL certifies five. For \textit{rdata}, the previously reported results for CP and HGTSA attain the corresponding best values on 22 of the 23 instances, whereas SAT proves all 23 optimal. These results suggest that increasing the number of factories substantially reduces the difficulty associated with machine-resource conflicts, despite enlarging the factory-assignment decision space.

\begin{figure}[htbp]
\centering
\resizebox{0.85\columnwidth}{!}{%
\begin{tikzpicture}
  \useasboundingbox (-3.1, -3.7) rectangle (3.1, 3.1);

  \definecolor{venngreen}{RGB}{130, 219, 130}   
  \definecolor{vennorange}{RGB}{254, 209, 126}  
  \definecolor{venngray}{RGB}{205, 205, 205}    
  \definecolor{vennpurple}{RGB}{198, 136, 194}  
  \definecolor{vennpink}{RGB}{223, 132, 212}    
  \definecolor{vennblue}{RGB}{121, 174, 232}    
  \definecolor{venncenter}{RGB}{180, 161, 181}  

  \def\circleA{(-0.95, 0.55) circle (1.85)}
  \def\circleB{(0.95, 0.55) circle (1.85)}
  \def\circleC{(0, -1.1) circle (1.85)}

  \begin{scope}[even odd rule]
    \clip \circleB (-5,-5) rectangle (5,5);
    \clip \circleC (-5,-5) rectangle (5,5);
    \fill[venngreen] \circleA;
  \end{scope}

  \begin{scope}[even odd rule]
    \clip \circleA (-5,-5) rectangle (5,5);
    \clip \circleC (-5,-5) rectangle (5,5);
    \fill[vennorange] \circleB;
  \end{scope}

  \begin{scope}[even odd rule]
    \clip \circleA (-5,-5) rectangle (5,5);
    \clip \circleB (-5,-5) rectangle (5,5);
    \fill[venngray] \circleC;
  \end{scope}

  \begin{scope}[even odd rule]
    \clip \circleC (-5,-5) rectangle (5,5);
    \clip \circleA;
    \fill[vennpurple] \circleB;
  \end{scope}

  \begin{scope}[even odd rule]
    \clip \circleB (-5,-5) rectangle (5,5);
    \clip \circleA;
    \fill[vennpink] \circleC;
  \end{scope}

  \begin{scope}[even odd rule]
    \clip \circleA (-5,-5) rectangle (5,5);
    \clip \circleB;
    \fill[vennblue] \circleC;
  \end{scope}

  \begin{scope}
    \clip \circleA;
    \clip \circleB;
    \fill[venncenter] \circleC;
  \end{scope}

  \node[font=\normalsize] at (-1.5, 2.7) {LBBD1\_CPIL};
  \node[font=\normalsize] at (1.5, 2.7) {CP};
  \node[font=\normalsize] at (0, -3.3) {SAT};

  \node[font=\normalsize] at (-1.55, 0.8) {4};
  \node[font=\normalsize] at (1.55, 0.8) {0};
  \node[font=\normalsize] at (0, -1.9) {3};
  \node[font=\normalsize] at (0, 1.4) {1};
  \node[font=\normalsize] at (-0.95, -0.45) {1};
  \node[font=\normalsize] at (0.95, -0.45) {35};
  \node[font=\normalsize] at (0, 0.1) {18};

\end{tikzpicture}
}
\caption{Venn diagram of instances proven optimal by LBBD1\_CPIL \cite{xiong2026lbBD}, CP \cite{meng2020milp, xiong2026lbBD}, and SAT across the two- and three-factory Brandimarte and \textit{rdata} configurations.}
\label{fig:venn_all_benchmarks}
\end{figure}
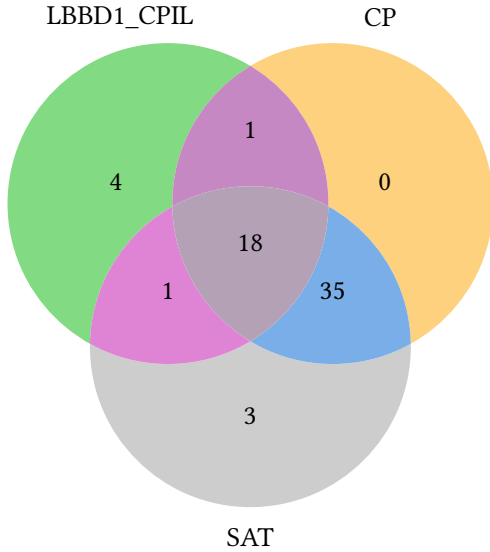

Figure~\ref{fig:venn_all_benchmarks} complements the configuration-level comparison above by showing how the optimality certificates of SAT, CP, and LBBD1\_CPIL overlap at the instance level across the Brandimarte and \textit{rdata} benchmarks. The comparison covers 66 instances from the two- and three-factory configurations, excluding Fattahi because corresponding CP and LBBD1\_CPIL results are unavailable.\footnote{For this figure, the CP set is the union of the original optimality results reported by \citet{meng2020milp} and the CP rerun reported by \citet{xiong2026lbBD}, with duplicate instances counted only once. Instances not evaluated by LBBD1\_CPIL remain in the 66-instance benchmark set but are not counted in the LBBD1\_CPIL set.} SAT extends the optimality coverage beyond that established by CP, proving four instances that are not certified by CP, including three SAT-only instances and one instance also certified by LBBD1\_CPIL. At the same time, the overlap between the two approaches is substantial, with 53 of the 54 instances proven optimal by CP also being certified by SAT. Thus, SAT adds new optimality certificates while covering nearly the entire set previously established by CP.

LBBD1\_CPIL provides further complementary optimality certificates. Five of its reported certificates lie outside the SAT set, including four unique to LBBD1\_CPIL and one shared with CP. Considering the union of the three exact approaches, optimality is established for 62 of the 66 Brandimarte and \textit{rdata} instances, leaving only four without a certificate among the available results. Together with the 10 Fattahi instances already proven optimal, 72 of the 76 benchmark instances considered in this study now have established optimal makespans.

\begin{table}[htbp]
\centering
\small
\setlength{\tabcolsep}{4.5pt}
\begin{threeparttable}
\caption{Benchmark instances for which SAT improves the previously best-known makespan or proves optimality.}
\label{tab:new_best_makespans}
\begin{tabular}{lccccc}
\toprule
\multirow{2}{*}{\textbf{Instance}} &
\multirow{2}{*}{\textbf{\#Factories}} &
\multirow{2}{*}{\textbf{BKV}} &
\multicolumn{2}{c}{\textbf{SAT}} &
\multirow{2}{*}{\textbf{Improvement}} \\
\cmidrule(lr){4-5}
& & & \textbf{Makespan} & \textbf{$t$ (s)} & \\
\midrule
MK10 & 2 & 130 & \textbf{125}$^{*}$ & $722.8$ & 5                  \\
MK08 & 3 & 184 & \textbf{184}$^{*}$ & $84.0$ & 0  \\
la15 & 3 & 388 & \textbf{387}$^{*}$ & $127.9$ & 1                  \\
\bottomrule
\end{tabular}
\begin{tablenotes}[flushleft]
\footnotesize
\item BKV denotes the best-known makespan previously reported in the literature.
\item An asterisk ($^{*}$) indicates that the corresponding makespan is proven optimal.
\end{tablenotes}
\end{threeparttable}
\end{table}

SAT also improves previously reported benchmark results, as summarized in Table~\ref{tab:new_best_makespans}. For \textit{MK10} with two factories, the previously best-known makespan of 130 is reduced to 125, and the improved value is proven optimal. For \textit{la15} with three factories, the best-known makespan is reduced from 388 to 387 and is likewise certified optimal. In addition, SAT proves that the previously reported makespan of 184 for \textit{MK08} with three factories is optimal. These results demonstrate that SAT can not only match existing benchmark results but also advance the state of the art by improving known bounds and establishing new optimality certificates.

The additional computational budget mainly affects the difficult two-factory cases. On Brandimarte, increasing the time limit from 600~s to 1800~s yields one additional optimality proof and one additional best makespan, including the improvement and certification of \textit{MK10}. For the two-factory \textit{rdata} configuration, the numbers of best and proven-optimal solutions remain unchanged, but the average deviation decreases from 0.39\% to 0.24\%, indicating that the longer runs improve incumbent solutions on some unresolved instances. Thus, the additional computational budget primarily strengthens solution quality and optimality certification on the remaining difficult cases rather than changing the overall performance pattern.

\section{Conclusion}
\label{sec:conclusion}

This paper has presented an exact SAT-based approach for the Distributed Flexible Job Shop Scheduling Problem (DFJSP) as a complementary alternative to existing exact methods. The proposed formulation jointly captures factory assignment, machine selection, operation timing, precedence relations, and conditional machine-capacity conflicts, while exploiting problem structure through temporal-domain reduction, propagation, symmetry breaking, and incremental SAT solving. The resulting framework enables both effective search for feasible schedules and rigorous certification of optimality, as demonstrated by the computational results on standard DFJSP benchmarks.

Computational results demonstrate the effectiveness of the proposed approach across 76 standard benchmark instances. Together with previously established results, the new SAT optimality certificates bring the total number of proven-optimal instances to 72 out of 76, including three newly certified instances. Moreover, SAT improves the previous best-known makespans for two of these instances: \textit{MK10} with two factories, from 130 to 125, and \textit{la15} with three factories, from 388 to 387, with both improved makespans proven optimal. For \textit{MK08} with three factories, SAT also proves the incumbent best-known makespan of 184 to be optimal. Across the benchmark configurations, SAT achieves the best results on four of the five configurations and remains competitive with strong exact approaches, including CP and LBBD1\_CPIL, as well as existing metaheuristic methods.

Future work will investigate stronger propagation mechanisms and more compact resource-conflict encodings for larger instances. Integrating SAT with heuristic warm-start strategies and tighter lower bounds is another promising direction. The framework can also be extended to more complex DFJSP variants, including sequencing flexibility, heterogeneous factories, and transportation times, and evaluated on additional benchmark instances.
\balance
\bibliographystyle{ACM-Reference-Format} 
\bibliography{references}

\end{document}